%% file: T.Komada_IHEP-seminar.tex
\documentclass[seceq]{ptptex}

\usepackage{graphicx}

\notypesetlogo                       %comment in if to eliminate PTPTeX 

\title{
Experimental Indication of Existence of \\
Extra Light-Vector Mesons $\omega'(1.3)$ and $\rho'(1.3)$ \\
}

\author{Toshihiko \textsc{\scshape Komada} }
\inst{%         %Affiliation, neglected when [addenda] or [errata]
      Department of Engineering Science, 
      Junior College Funabashi Campus, \\
      Nihon University, Funabashi 274-8501, Japan 
}
\abst{%         %this abstract is neglected when [addenda] or [errata]
Mass distributions of $\pi^+ \pi^- \pi^0$ and $\omega \pi^0$ systems 
in the $e^+e^-$ annihilation 
obtained by the several experiments 
are analyzed. We obtain indication of 
light-vector mesons 
$\omega'(1.3)$ and $\rho'(1.3)$ 
in the lower mass region.
Those  states are expected to exist in the $\widetilde U(12)$-classification scheme.
}
\begin{document}

\maketitle

%%%%%%%%%%%%%%%%%%%%%%%%%%%%%%%%%%%%%%%%%%%%
%% MAINMATTER
\hfill
%%%%%%%%%%%%%%%%%%%%%%%%%%%%%%%%%%%%%%%%%%%%

\section{  Introduction  }
Recently the $\widetilde U(12)$ level-classification scheme of hadrons\cite{U12} 
has been proposed, 
which is a covariant generalization of 
non-relativistic scheme based on $SU(6)_{SF}$. 
In the $\widetilde U(12)$-classification scheme 
a new type of relativistic states, 
 called chiral states, 
 which have no correspondents in the non-relativistic scheme, 
are expected to exist in lower mass region for light quark systems. 
 Especially 
the extra vector meson nonet 
is predicted 
 in the ground S-wave state of ($q \bar q$) system.

 On the other hand, experimentally it 
is noted that the recent data of $3 \pi$ and $\omega \pi^0$ states 
in $e^+e^-$ annihilation show a hint for 
an extra state $\omega(1200)$\cite{SND:omega}. 
The studies in hadroproduction and others 
have shown indication of low mass extra states\cite{donnachie}. 
 Further studies on their existence will be interesting.

 In this work we 
re-analyzed  
mass spectra of 
$\pi^+ \pi^- \pi^0$ \ \cite{SND03:omega,BABAR} and 
$\omega \pi^0$ \ \cite{CMD-2:rho,SND:rho,DM2:rho} 
in $e^+e^-$ annihilation, 
and have obtained indication of a low mass $\omega$ 
and a low mass $\rho$ in respective channels around 1.3 GeV. 
These seem to be ground-state chiralons  (pure chiral state)  
expected  in the $\widetilde U(12)$ scheme\cite{yamada:proc}. 

% --------------------------------------------
\section{
Analysis of the mass spectrum of $\pi^+ \pi^- \pi^0 $ and 
indication of ~~~~ extra vector meson $\omega'(1.3)$}

\begin{figure}[hbpt]
 \begin{center}
  \includegraphics[width=10cm,angle=0]{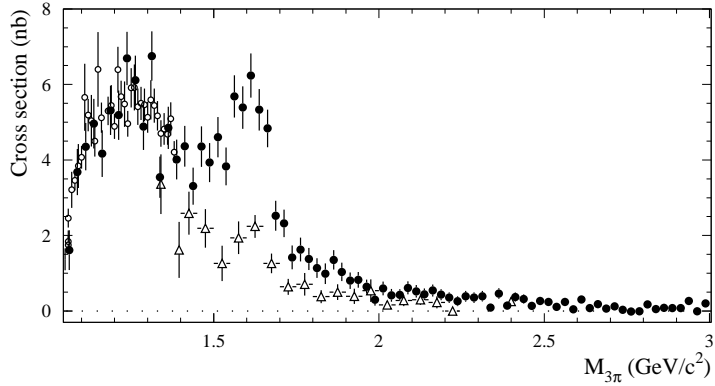}
  \caption{
The $e^{+}e^{-}\rightarrow\pi^{+}\pi^{-}\pi^{0}$ cross section 
by BABAR\cite{BABAR} (filled circles), 
by SND\cite{SND03:omega} (open circles), and by DM2\cite{DM2:rho} (open triangles).
}
  \label{fig:BABAR}
 \end{center}
\end{figure}

In this work we are going to reanalyze the combined mass spectrum of the  $\pi^{+}\pi^{-}\pi^{0}$ data in the $e^{+}e^{-}$ annihilation obtained by SND\cite{SND03:omega}  and by BABAR\cite{BABAR}.  
The former presents higher statistics data at the lower mass region and the latter covers the whole mass region interested as shown in Fig.  \ref{fig:BABAR}. 
The DM2 data\cite{DM2:omega} which were included in our previous analysis\cite{komada:Had05} are not used in the present analysis since they seem to show different behaviors from those of BABAR depending on a bias factor for cross sections.

The relevant process $e^+ e^- \to \pi^{+}\pi^{-}\pi^{0}$ is, 
applying the vector meson dominance   
model (VMD), considered to 
occur dominantly through intermediate production of   
vector mesons ``V'' as 
that $e^+ e^- \to \gamma \to ``V" \to \rho \pi \to 3 \pi$.  
The analysis  results obtained by SND and by BABAR show vectors,
$\omega(782),\ \phi(1020),\ \omega(1420)$ and $\omega(1650)$, but no $ \omega(1200)$. The width parameter for $\omega(1420)$ is obtained to be rather wider in each analysis, as shown in Table I. 
There recognized, however, a huge event accumulation exists around 1.3 GeV 
in the $ 3\pi$ mass spectrum. 
It may be 
 naturally interpreted   to correspond to the $\omega(1200)$, 
which was  pointed out in ref. \citen{SND:omega}, 
 rather than $\omega(1420)$.

In order to make clear the situation on the existence of 
$ \omega(1200)$,   
a possible contribution of a low mass state 
$\omega'(1.3)$   is considered explicitly 
in the present work in addition to the higher vector mesons, 
$\omega(1420)$ and $\omega(1650)$.

%%%%%%%%%%%%%%%%%%%%%%%%%%%%%%%%%%%%%%%%%
%%%%%%%%%%%%%%%%%%%%%%%%%%%%%%%%%%%%%%%%%%

\begin{table}[bhpt]
\begin{center}
\caption{Masses and widths of $\omega(1420)$ in $e^{+}e^{-}\rightarrow\pi^{+}\pi^{-}\pi^{0}$ are listed.}
\label{tab1}
\begin{tabular}{|c|c|c|}
\hline
$\omega (1420)$ & M (MeV) & $\Gamma$ (MeV) \\ \hline

BABAR\cite{BABAR}  & $1350 \pm 20 \pm 20$  & $450 \pm 70 \pm 70$  \\ \hline

SND\cite{SND03:omega}  & $1400 \pm 50 \pm 130$ & $870 ^{+500}_{-300} \pm 450 $  \\ \hline

 \end{tabular}
\end{center}
\end{table}
%
%
%-----------------------------------

\subsection{Method of analysis}
First we give general 
effective Lagrangians, 
which concern on our relevant processes given in Fig. 2, as 
\begin{eqnarray}
{\cal L}_{e^+e^-\gamma }
&=&ie\bar \psi _e\gamma _\mu \psi _e{\bf A_\mu }, \nonumber \\
{\cal L}_{\gamma V}
&=&(
\frac{e m_\omega ^2}{3f_\omega }\omega_\mu  -  
\frac{\sqrt{2}em_\phi ^2}
{3f_\phi }\phi_\mu)A_\mu,  \nonumber \\
 & & (``V" = \omega(782) , \phi(1020) , \omega^{(1)}, \omega^{(2)} \ \mbox{\rm and } \ \omega^{(3)} ), 
\nonumber \\
{\cal L}_{\omega \rho \pi } &=& 
g_{\omega \rho \pi } 
\varepsilon _{\mu \nu \lambda \kappa }
\partial _\mu \omega _\nu \partial _\lambda
\rho_\kappa \cdot{ \pi  } , 
\nonumber \\
{\cal L}_{\phi  \rho \pi } &=& 
g_{\phi  \rho \pi }
\varepsilon _{\mu \nu \lambda \kappa }
\partial _\mu \phi  _\nu \partial _\lambda
\rho_\kappa \cdot{\pi } , 
\nonumber \\
{\cal L}_{\rho \pi\pi } &=& 
-f_{\rho \pi \pi }\rho _\mu 
(\pi \times \partial _\mu \pi  ),  
\end{eqnarray}
%
%
%
%
%%%%%%%%%%%%%%%%%%%%%%%%%%%%%%%%%%%%%%%%%%%%%%%%%%%%%%%%%
where $f_V$ being coupling constant of decay interaction for $ ``V" \rightarrow e^{+}e^{-}$.
The cross section for relevant process is given as 
%%%%%%%%%%%%%%%%%%%%%%%%%%%%%%%%%%%%%%%%%%%%%%%%%%%%%%%%%
\begin{eqnarray}
&\sigma (s)
 =& \frac{4\pi \alpha ^2}{s^{\frac{3}{2}}}
\nonumber \\
 & & \left|  
     A_\omega  
    \frac{m_\omega ^2\sqrt{F_\omega(s)}}{m_\omega ^2-s-im_\omega \Gamma _\omega }
  + A_\phi    \frac{m_\phi ^2\sqrt{F_\phi(s)}}{m_\phi ^2-s-im_\phi \Gamma _\phi }
  + \sum_{i=1}^3 A_i\frac{m_{\omega^{(i)} } ^2\sqrt{F_{\omega^{(i)} }(s)}}{m_{ \omega^{(i)} }^2-s-im_{\omega^{(i)} }\Gamma _{\omega^{(i)}}} 
\right|^24\pi \Gamma (s), \nonumber \\
%%%%%%%%%%%%%%%%%%%%%%%%%%
\end{eqnarray}
where 
$\omega^{(1)}, \omega^{(2)}$ and $\omega^{(3)} $ denote 
a low mass and two higher mass vector states 
above 1 GeV.

%%%%%%%%%%%%%%
In Eq. (2.2) we have introduced the form factor $F$ given as 
\\
%%%%%%%%%%%%%%%%%%%%%%%%%%
\begin{eqnarray}
F_R (s) = \frac{2m_R^2}{m_R^2+s} , 
\end{eqnarray}
%%%%%%%%%%%%%%%%%%%%%%%%%%
and $A_i$'s are the fitting parameters, 
 while the values of $A_\omega$ and $A_\phi$ has been estimated 
 from the relevant low mass data. 
%%%%%%%%%%%%%%%%%%%%%%%%%%
\begin{eqnarray}
\Gamma(s)
&=& \frac{f_{\rho \pi \pi }^2}{3\cdot 16\pi ^3}\int_{2m_\pi }^{\sqrt{s}-m_\pi }d\sqrt{s_{12}}{\bf |q|^3}{\bf |p_1|^3}\cdot \int_{-1}^1d(cos\theta )sin^2\theta 
 \left| {\cal F}_{\rho \pi \pi}
   \right|^2 , \nonumber \\
{\cal F}_{\rho \pi \pi} &=&
 \frac{\sqrt{F_\rho (s_{12})}}{m_\rho ^2-s_{12}-im_\rho \Gamma _\rho }
     +\frac{\sqrt{F_\rho (s_{23})}}{m_\rho ^2-s_{23}-im_\rho \Gamma _\rho } +\frac{\sqrt{F_\rho (s_{31})}}{m_\rho ^2-s_{31}-im_\rho \Gamma _\rho }, \nonumber \\
 &&( f_{\rho \pi \pi }^2  = ``1"), \hfill
\end{eqnarray}
%%%%%%%%%%%%%%%%%%%%%%%%%%
%%%%%%%%%%%%%%
where $s_{ij}$'s are invariant mass of three possible contribution 
of $\pi^+ ~,~ \pi^-$ and $\pi^0$.
$\theta$ is angle between $\rho$ and $\pi$ in $\omega $ at rest. 
%%%%%%%%%%%%%%
${\bf q}$ and ${\bf p_1}$ is 3-momentum of 
${\bf p_{\pi_3}}$ at $\pi_1\pi_2\pi_3$ C.M.S. and of 
${\bf p_{\pi_1}}$ at $\pi_1\pi_2$ C.M.S., respectively.
%%%%%%%%%%%%%%
The value of common factor of the cross section, $f_{\rho \pi \pi }^2$, 
is irrelevant to our analysis.  
%%%%%%%%%%%%%%

There are two contrastive 
structures 
in $3\pi$ mass spectrum of $e^{+}e^{-}\rightarrow\pi^{+}\pi^{-}\pi^{0}$ below 2 GeV. 
One is below 1 GeV region and the other is above 1 GeV region. 
%
%
%
%%%%%%%%%%%%%%%%%%%%%%%%%%%%%%%%%%%%%%%%%
\begin{figure}[hbpt]
\begin{center}
\input{./fig-diagram_omega.tex}
  \caption{
Diagram of $e^{+}e^{-}\rightarrow\gamma\rightarrow\pi^{+}\pi^{-}\pi^{0}$.
}
  \label{fig:diagram-omg}
\end{center}
\end{figure}
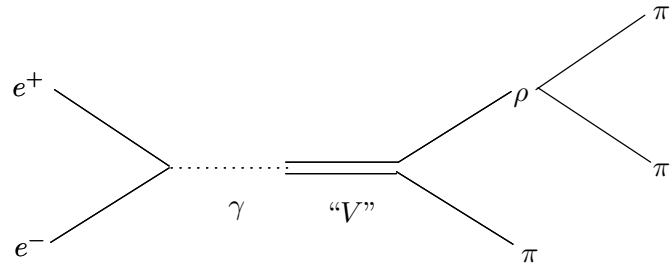
%%%%%%%%%%%%%%%%%%%%%%%%%%%%%%%%%%%%%%%%%
The former region has 
 two clear and huge peaks 
coming from contributions from $\omega(782)$ and $\phi(1020)$
 as shown in Fig. \ref{fig:SND}a),    
while the latter shows some complex structures   
(which 
 are relevant to the present work) 
as shown in Fig. \ref{fig:BABAR} and  Fig. \ref{fig:SND}b).

%%%%%%%%%%%%%%%%%%%%%%%%%%%%%%%%%%%%%%%%%
\begin{figure}[hbpt]
 \begin{center}
  \includegraphics[width=14cm,angle=0]{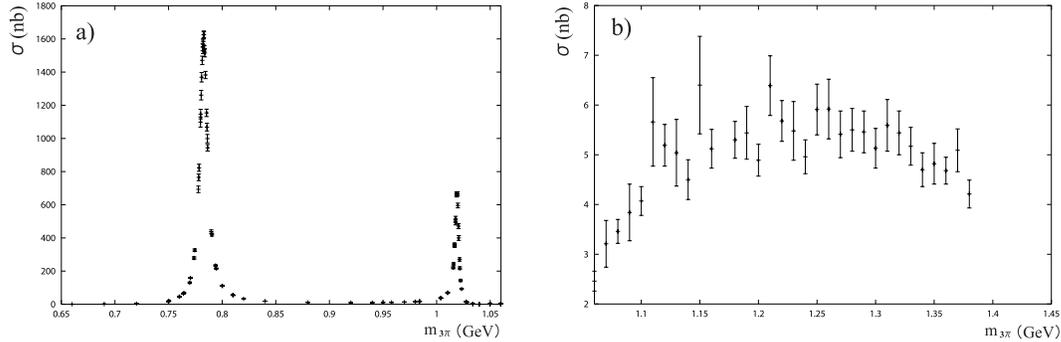}
  \caption{
The $e^{+}e^{-}\rightarrow\pi^{+}\pi^{-}\pi^{0}$ cross section 
by SND\cite{SND03:omega}, 
a) below 1 GeV region c) above 1 GeV region.
}
  \label{fig:SND}
 \end{center}
\end{figure}
%%%%%%%%%%%%%%%%%%%%%%%%%%%%%%%%%%%%%%%%%%
Before $3 \pi$ mass spectrum of $e^{+}e^{-}\rightarrow\pi^{+}\pi^{-}\pi^{0}$ 
in the energy region from 1.06 to 2.0 GeV by BABAR and by SND are analyzed, 
the parameters
$A_\omega $ and $A_\phi$ 
of $\omega(782)$ and $\phi(1020)$ in Eq. (2.2)  
are fixed by the analysis of the spectrum below 1 GeV. 
Then these values\footnote{
$A_\omega= 0.64 $ , $A_\phi=-0.041$.}  
of  parameter $A_\omega $ and $A_\phi$ are 
applied in the analysis of the data 
above  
1 GeV.  
% ----------------
%
%
%

% 
%
%-----------------------------------

\subsection{Results of analysis}
 The $3 \pi$ mass spectrum of $e^{+}e^{-}\rightarrow\pi^{+}\pi^{-}\pi^{0}$ 
in the energy region from  1.06 to 2.0   GeV are 
used in the analyses of the following three cases. 

In the fitting of the  first case, two resonances are considered above 1 GeV region.
The values  of mass and width of 
$\omega^{(1)}$ and $\omega^{(2)}$, corresponding to $\omega(1420)$ and $\omega(1650)$,  
are restricted to be  consistent with PDG tables\cite{PDG2004}.
 In this case, the experimental data are not well reproduced, 
especially below 1.5 GeV region.

 In the  second case, two resonances are considered above 1 GeV region.
The values of mass and width of 
$\omega^{(2)}$, corresponding to $\omega(1650)$,  
are restricted 
to be   
consistent with PDG tables\cite{PDG2004}, 
while the parameters of 
$\omega^{(1)}$ 
are not restricted.
 In this case,  the experimental data are well reproduced, 
although obtained values of mass and width of $\omega^{(1)}$ 
are lower and wider, respectively, compared with $\omega(1420)$  
of PDG values.

 In the third case, three resonances are considered above 1 GeV region.
The values  of mass and width of 
$\omega^{(2)}$ and $\omega^{(3)}$, corresponding to $\omega(1420)$ and $\omega(1650)$,  
are restricted to be  consistent with PDG tables\cite{PDG2004}, 
while the parameters of $\omega^{(1)}$, 
supposed  to be corresponding to the extra $\omega'(1.3)$,  are not restricted.
 In this case, the experimental data are well reproduced. 
The  contributions of $\omega^{(1)}$ and $\omega^{(3)}$  are large, while the 
contribution of $\omega^{(2)}$ is very small.

\begin{figure}[hbpt]
\begin{center}
  \includegraphics[width=7.5cm,angle=0]{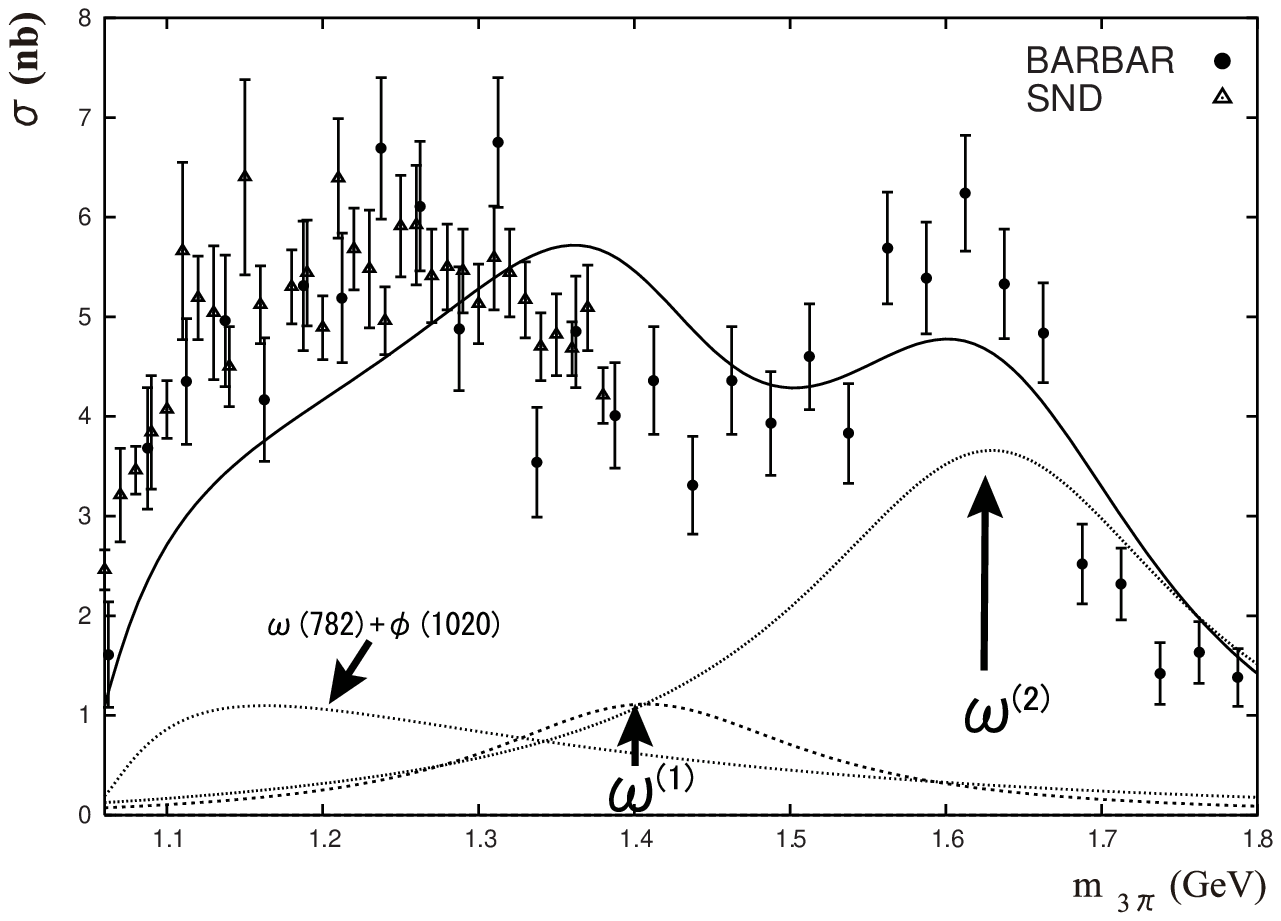}
  \caption{
Results of the analysis of the 
first case (fit with $ \omega^{(1)} ,\ \omega^{(2)}$) 
on $3 \pi$ mass spectrum of $e^{+}e^{-} \rightarrow \pi^+ \pi^- \pi^0 $. 
Data are from SND\cite{SND03:omega}and BABAR\cite{BABAR}. 
Solid line is  fitted curve.  
Dotted lines  represent the contribution of  each amplitude.
}
  \label{fig:omg1}
\end{center}
\end{figure}

\begin{figure}[hbpt]
\begin{center}
  \includegraphics[width=7.5cm,angle=0]{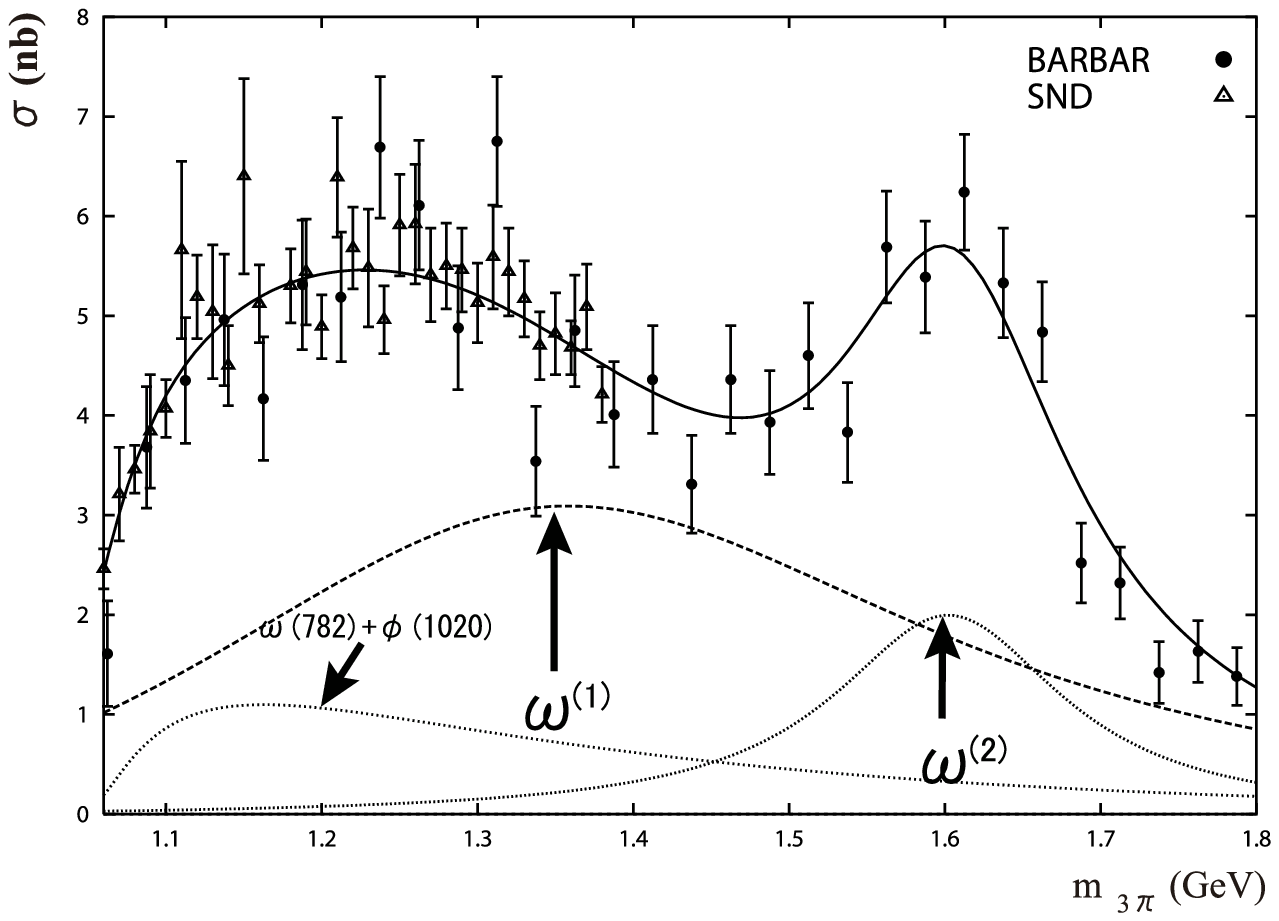}
  \caption{
Results of the analysis 
    of 
the 
second case (fit with $ \omega^{(1)} ,\ \omega^{(2)}$)
on $3 \pi$ mass spectrum of $e^{+}e^{-} \rightarrow \pi^+ \pi^- \pi^0 $. 
   Data are from SND\cite{SND03:omega}and BABAR\cite{BABAR}. 
Solid line is  fitted curve.  
Dotted lines  represent the contribution of  each amplitude.
}
  \label{fig:omg2}
\end{center}
\end{figure}

\begin{figure}[hbpt]
\begin{center}
\includegraphics[width=7.5cm,angle=0]{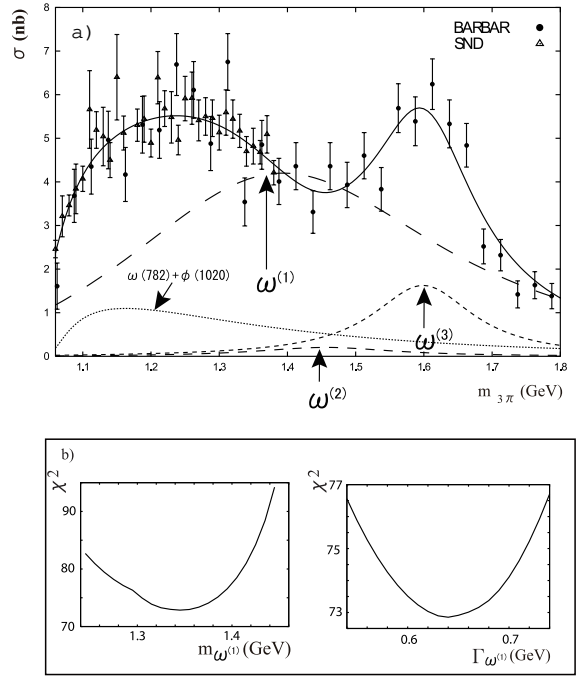}
  \caption{
a) Results of the analysis 
    of 
the 
third case (fit with $ \omega^{(1)} ,\ \omega^{(2)} ,\ \omega^{(3)}$) 
on $3 \pi$ mass spectrum of $e^{+}e^{-} \rightarrow \pi^+ \pi^- \pi^0 $. 
b) Mass and width scan on $\omega^{(1)}$ for 
the 
third case.
   Data are from SND\cite{SND03:omega}and BABAR\cite{BABAR}. 
Solid line is  fitted curve.  
Dotted lines  represent the contribution of  each amplitude.
}
  \label{fig:omg3}
\end{center}
\end{figure}

 All the results of three cases 
 are shown in Fig. \ref{fig:omg1}, Fig. \ref{fig:omg2}, and Fig. \ref{fig:omg3}a).
 In addition mass  and width scan on $\omega^{(1)}$ for 
the third case are shown in Fig. \ref{fig:omg3}b).
The obtained values of parameters and $\chi^2 / N_{d.o.f.}$ 
 in the three cases  are listed in Table II.

The almost same values of $\chi^2 / N_{d.o.f.}$ are obtained in 
the second and the third case. 
This reflect that the contribution of $\omega(1420)$ in the third case 
is very small, and implies that a dominant contribution of 
$\omega'(1.3)$ in this case is replaced by that of $\omega(1420)$ 
in the second case.

The result of fitting of the third case   
is improved by about $19\sigma$ 
 compared with   the first case, indicating the existence of 
the low mass extra vector meson $\omega'(1.3)$ in addition to the two 
higher states 
$\omega(1420)$ and $\omega(1650)$.

It is noted that the obtained values of $m_{\omega^{(1)}}$ in 
the second and third case are slightly larger than 
the center of accumulation, around 1.25 GeV, in $3 \pi$ mass spectrum 
as shown in Fig. \ref{fig:BABAR} and  Fig. \ref{fig:SND}b).  
It is due to interference effect among $\omega(782)$, $\phi(1020)$, and $\omega'(1.3)$.

\vspace{-1em}

\begin{table}[hbpt]
\begin{center}
\caption{
The obtained values of parameters and $\chi^2 / N_{d.o.f.}$.
}
\begin{tabular}{|c|c|c|c|c|c|c|}
\hline
\multicolumn{7}{|c|}{ Two resonance analysis  } \\   \hline
        &     \multicolumn{3}{|c|}{ ~~~~~~~~~~~~   $\omega^{(1)}$  ~~~~~~~~~~~~ }  &
              \multicolumn{3}{|c|}{ $\omega^{(2)}$ }  \\\hline
Case 1  & \multicolumn{6}{c|}{ 
$ \chi^2/N_{d.o.f.}  = 442/64 = 6.91 $  
} \\ 
$m(\mbox{\rm MeV})$        &  \multicolumn{3}{|c|}{ $1400^\dagger $  }  &
                              \multicolumn{3}{|c|}{ $1626 \pm 5$ }  \\

$\Gamma (\mbox{\rm MeV})$  &  \multicolumn{3}{|c|}{ $250^\dagger$ }  
                           &  \multicolumn{3}{|c|}{ 280 $^\dagger$ }  \\

 A                         &  \multicolumn{3}{|c|}{ $ -0.039\pm0.002 $  }  &
                              \multicolumn{3}{|c|}{ $-0.058 \pm 0.002 $  }  \\ 

\hline % -----------------------------------------------------------------

Case 2  & \multicolumn{6}{c|}{ 
$ \chi^2/N_{d.o.f.}  = 76/64=1.19$
  } \\ 
$m(\mbox{\rm MeV})$        &  \multicolumn{3}{|c|}{ 1303 $\pm 10 $  }  &
                              \multicolumn{3}{|c|}{ 1601 $\pm 5$ }  \\

$\Gamma (\mbox{\rm MeV})$  &  \multicolumn{3}{|c|}{ $ 641 \pm 49 $ }  
                           &  \multicolumn{3}{|c|}{ 180$^\ddagger$  }  \\

 A                         &  \multicolumn{3}{|c|}{ $-0.191 \pm 0.018 $  }  &
                              \multicolumn{3}{|c|}{ $ -0.027 \pm 0.003 $   }  \\ 

\hline \hline
%%%%%%%%%%%%%%%%%%%%%%%
\multicolumn{7}{|c|}{ Three resonance analysis  } \\   \hline

                           &  \multicolumn{2}{|c|}{ $\omega^{(1)}$  }  &
                              \multicolumn{2}{|c|}{ $\omega^{(2)}$   }  &
                              \multicolumn{2}{|c|}{ $\omega^{(3)}$ }  \\ \hline

Case 3  & \multicolumn{6}{c|}{ 
$ \chi^2/N_{d.o.f.}  = 73/61=1.19$  
} \\ 
$m(\mbox{\rm MeV})$        &  \multicolumn{2}{|c|}{ 1346 $\pm 26$  }  &
                              \multicolumn{2}{|c|}{ 1450$^\dagger$   }  &
                              \multicolumn{2}{|c|}{ 1597 $\pm 6$ }  \\ 

$\Gamma (\mbox{\rm MeV})$  &  \multicolumn{2}{|c|}{ $639 \pm 42 $ }  &
                              \multicolumn{2}{|c|}{ $ 250 ^\dagger $ }  &
                              \multicolumn{2}{|c|}{ $180 ^\ddagger $ }  \\ 

 A                         &  \multicolumn{2}{|c|}{ $-0.207 \pm 0.015  $  }  &
                              \multicolumn{2}{|c|}{ $ 0.015 \pm 0.010 $   }  &
                              \multicolumn{2}{|c|}{ $-0.026 \pm 0.003$  }  \\ \hline
 \multicolumn{7}{l}{ 
$^\dagger$ Bound for upper limit which is set to be consistent 
with PDG tables. } \\
 \multicolumn{7}{l}{ 
$^\ddagger$ Bound for lower limit which is set to be consistent 
with PDG tables. } 
 \end{tabular}
\end{center}
\end{table}
%%%%%%%%%%%%%%%%%%%%%%%%%%%%%%%%%%%%%%%%%%%%%%%%%%%%%%%%%%%%%%%
\vspace{-1em}
%%%%%%%%%%%%%%%%%%%%%%%%%%%%%%%%%%%%%%%%%

\vspace{-1em}

 \newpage

%%%%%%%%%%%%%%%%%%%%%%%%%%%%%%%%%%%%%%%%%
%%%%%%%%%%%%%%%%%%%%%%%%%%%%%%%%%%%%%%%%%
%-----------------------------------

 \ \ \ \\
 \newpage

% =================================================================

\section{
Analysis of the mass spectrum of $\omega \pi^0$ and 
indication of ~~~~~~~~~ extra vector meson $\rho'(1.3)$}

In this work we are also going to reanalyze  
the combined mass spectrum of $\omega \pi^0 $ data 
in $e^+e^-$ annihilation 
obtained by CMD-2\cite{CMD-2:rho}, 
by  SND\cite{SND:rho} and 
by  DM2\cite{DM2:rho}. 
The CLEO data\cite{CLEO:rho} which were included 
in our previous analysis\cite{komada:Had05} are not used in the present analysis 
to make the situation simple for analysis 
since the CLEO data were obtained from $\tau \to \omega \pi^0$ 
which is the different process from  $e^+ e^- \to \omega \pi^0$.

The process is also, applying the VMD, 
considered to occur dominantly through  
intermediate production of 
vector mesons ``V'' as that  
$e^+ e^- \to \gamma \to ``V" \to \omega \pi^0$.    
The analysis  results obtained by CMD-2 show vectors,  
$\rho(770), ~\rho(1450)$ and $\rho(1700)$, 
while mass and width values of the $\rho(1450)$ are 
scattered in each study\cite{PDG2004}. 
There   
recognized, however, a huge event accumulation exists around 1.3 GeV   in the $\omega \pi^0$  mass spectrum.    
It may be naturally 
 interpreted to correspond to the 
$\rho'(1.3)$ with mass around 1.3 GeV which has lower mass than $\rho(1450)$.   
Actually the existence of $\rho'(1.3)$  was pointed out 
by several experimental group\cite{LASS:rho,OBELIX:rho} and 
in review articles\cite{PDG2004,donnachie}.

In order to make clear the situation on the existence of 
extra light-vector meson,   
a possible contribution of 
$\rho'(1.3)$ is considered explicitly in the present work   
in addition to the higher vector mesons, $\rho(1450)$ and $\rho(1700)$.
%
%%%%%%%%%%%%%%%%%%%%%%%%%%%%%%%%%%%%%%

\subsection{Method of analysis}

Effective Lagrangians, 
which concern on our relevant processes given in Fig. 7, 
 are given as 
%
%%%%%%%%%%%%%%%%%%%%%%%%%%%%%%%%
\begin{eqnarray}
{\cal L}_{e^+e^- \gamma} & = & i e \bar \psi_e \gamma_\mu \psi_e A_\mu,  \nonumber \\
%%%%%%%%%
{\cal L}_{\gamma_V}  & = &  \gamma_{V} V_\mu^3A_\mu, 
               (\gamma_V=\frac{em_V^2}{f_V})        \nonumber  \\
%%%%%%%%%
{\cal L}_{V \omega \pi}  & = & 
                g_{V \omega \pi}
                \varepsilon _{\mu \nu \lambda \kappa}
                \partial_\mu {\bf V}_\nu 
                \partial_\lambda 
                \omega_\kappa{\bf \pi}  \nonumber \\
 & & (``V" = \rho(770) ,  \ \rho^{(1)}, \  \rho^{(2)}, \ \mbox{\rm and } \ \rho^{(3)} ),  \label{eqL5} 
\end{eqnarray}
%%%%%%%%%%%%%%%%%%%%%%%%%%%%%%%%
%
where $f_V$ being coupling constant of decay interaction for $ ``V" \rightarrow e^{+}e^{-}$.
$\rho^{(1)},~ \rho^{(2)}$ and $\rho^{(3)} $ denote a low mass and two higher mass 
vector states. 
The cross section is given as \\
\begin{eqnarray}
{\sigma}_{0}(s)
=
\frac{4 \pi {\alpha}^2}{s^{\frac{3}{2}}}
{(\frac{g_{\rho \omega \pi}}{f_{\rho}})}^2
\left|
\frac{m_{\rho}^2 \sqrt{F_\rho(s)}}
{D_{\rho}(s)}
+ 
\sum_{i=1}^3 
A_{i}
\frac{m_{\rho^{(i)} }^2 \sqrt{F_{\rho^{(i)}}(s)}}
{
m_{\rho^{(i)}}^2 -s -i m_{\rho^{(i)}} \Gamma_{\rho^{(i)}}
}  
\right|^2
P_f(s) ,  
\end{eqnarray}
where 
\begin{eqnarray}
%%%%%%%
F_R(s) = \frac{2m_R^2}{m_R^2+s}  \ \ , \ \ 
P_f(s)=\frac{1}{3}{\left|{\bf p}_{\omega}(s)\right|}^3 \cdot B_{\omega \rightarrow \pi ^0 \gamma } ~~.  %\nonumber
\end{eqnarray}
${\bf p}_{\omega}(s)$ is three momentum of $\omega$ in $\rho$ at rest. 
$B_{\omega \rightarrow \pi ^0 \gamma }$ is branching ratio of $\omega \rightarrow \pi ^0 \gamma$ to be $0.085 \pm 0.005$. 
%
%
%%%%%%%%%%%%%%
Coupling constants 
are estimated by VMD using experimental values to be 
$f_\rho = 5.04$ and $g_{\rho \omega \pi } =12.47$.~$A_i$'s are the fitting parameters.
%
%%%%%%%%%%%%%%%
%
%
%%%%%%%%%%%%%%%%%%%%%%%%%%%%%%%%%%%%%%%%%
\begin{figure}[hbpt]
\begin{center}
   \includegraphics[width=8cm,angle=0]{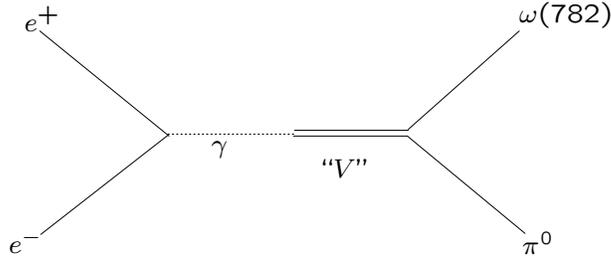}
  \caption{
Diagram of $e^{+}e^{-}\rightarrow\gamma\rightarrow\omega\pi^{0}$.
}
  \label{fig:diagram-rho}
\end{center}
\end{figure}
%%%%%%%%%%%%%%%%%%%%%%%%%%%%%%%%%%%%%%%%%
% --------------------------------------------
\subsection{Results of analysis}

The $\omega \pi$ mass spectrum of $e^{+}e^{-} \rightarrow \omega \pi^{0}$ 
below 2.2 GeV are used in the analyses of following three cases 
in parallel with the analyses
on $\pi^+ \pi^- \pi^0$.  
%-----------------------------------

In the fitting of the  first case, two resonances are 
considered above 1 GeV region.
%-----------------------------------
The values   of mass and width of 
$\rho^{(1)}$ and $\rho^{(2)}$, corresponding to 
$\rho(1450)$ and $\rho(1700)$,   
are restricted to be consistent with PDG tables\cite{PDG2004}. 
%-----------------------------------
In this case, the experimental data are not well reproduced, 
especially below 1.4 GeV region.
%-----------------------------------

In the second case, 
two resonances are considered above 1 GeV region.
%-----------------------------------
The values   of mass and width of 
$\rho^{(2)}$, corresponding to $\rho(1700)$,    
are restricted to be consistent with PDG tables\cite{PDG2004}, 
while the parameters of $\rho^{(1)}$ are not restricted.
%-----------------------------------
In this case, 
the experimental data are well reproduced, although obtained values of 
mass and width of $\rho^{(1)}$ are lower and wider, respectively, 
compared withthose of  $\rho(1450)$ of PDG values. 
%-----------------------------------

In the third case, 
three resonances are considered above 1 GeV region.
%-----------------------------------
The values  of mass and width of 
$\rho^{(2)}$ and $\rho^{(3)}$, corresponding to 
$\rho(1450)$ and $\rho(1700)$,    
are restricted to be consistent with PDG tables\cite{PDG2004}, 
respectively, while the parameters of 
$\rho^{(1)}$ , supposed to be corresponding to the extra $\rho'(1.3)$,  are not restricted.
%-----------------------------------
In this case, 
the experimental 
data 
are 
well reproduced. 
The contribution from 
$\rho^{(1)}$ is large, 
while those of $\rho^{(2)}$ and $\rho^{(3)}$ are very small 
comparing to that of $\rho^{(1)}$. 
%-----------------------------------

\begin{figure}
\begin{center}
  \includegraphics[width=7.5cm,angle=0]{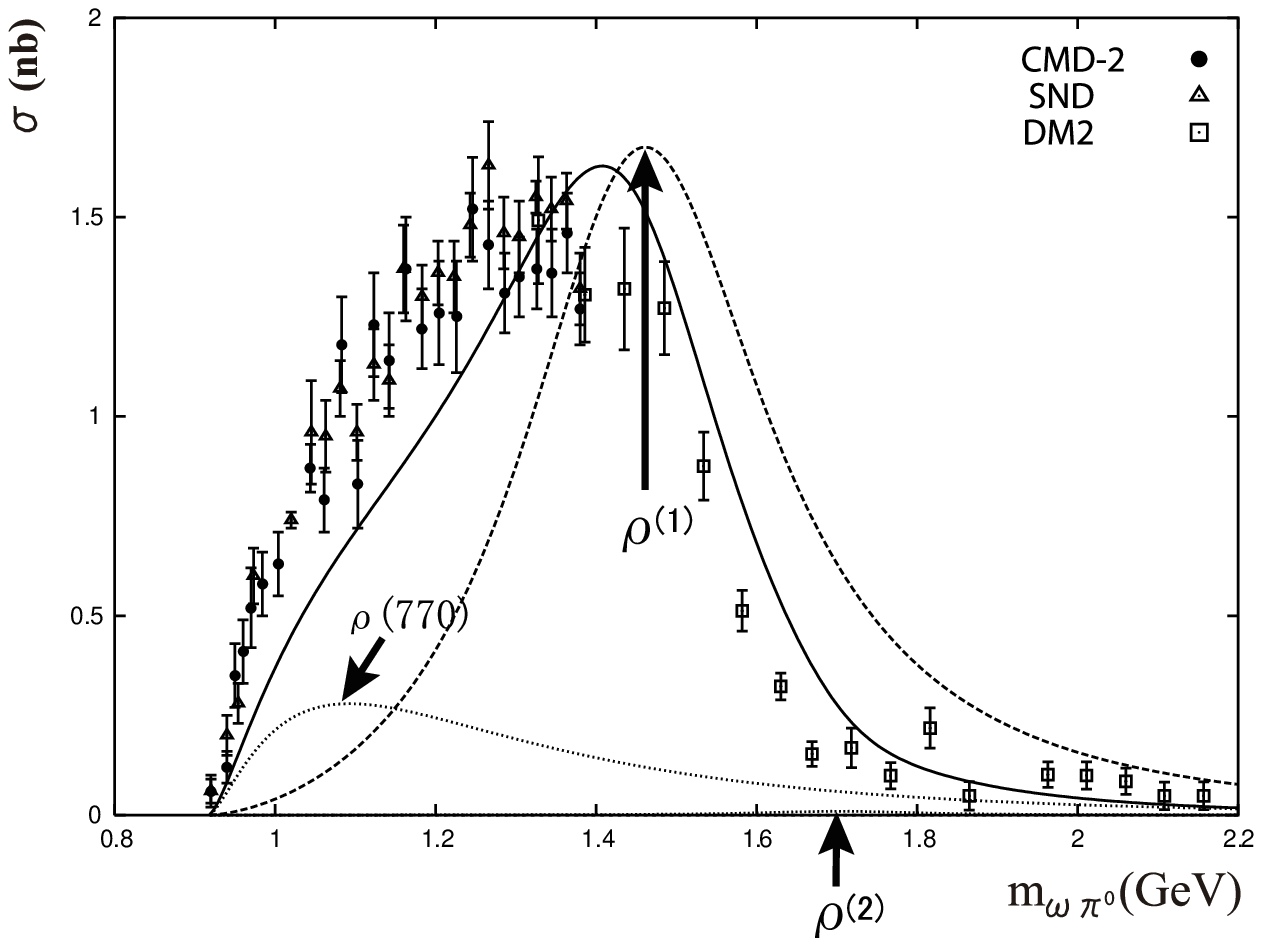}
  \caption{
Results of the analysis 
    of 
the 
first case (fit with 
$ \rho^{(1)} ,\ \rho^{(2)}$ 
) 
on $\omega \pi^0$ mass spectrum of $e^{+}e^{-} \rightarrow \omega \pi^0 $.
Data are from CMD-2\cite{CMD-2:rho}, SND\cite{SND:rho}, and 
DM2\cite{DM2:rho}. 
Solid line is  fitted curve.  
Dotted lines  represent the contribution of  each amplitude.
}
  \label{fig:res-rho1}
\end{center}
\end{figure}

\begin{figure}
\begin{center}
  \includegraphics[width=7.5cm,angle=0]{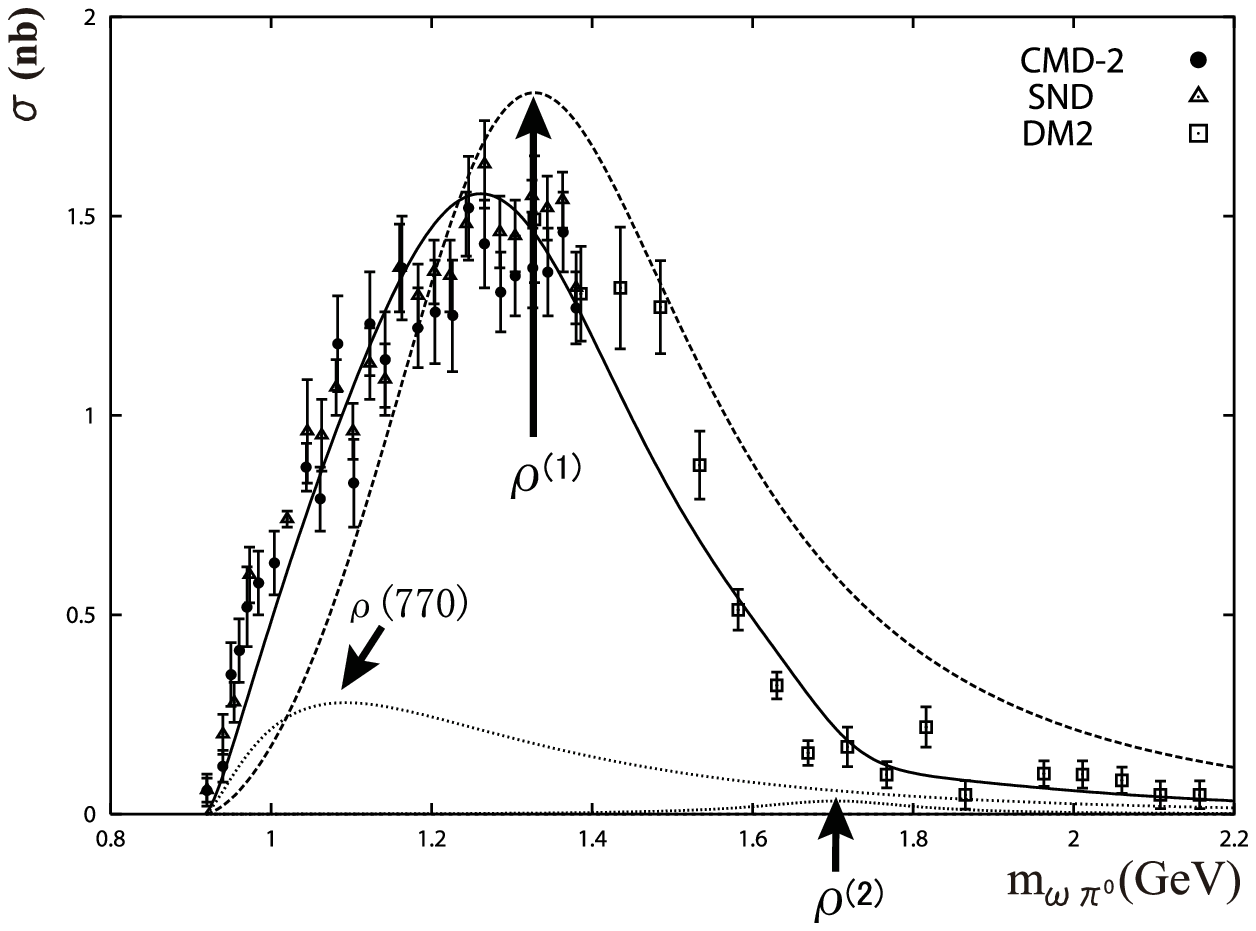}
  \caption{
Results of the analysis 
    of 
the 
second case (fit with 
$ \rho^{(1)} ,\ \rho^{(2)}$ 
) 
on $\omega \pi^0$ mass spectrum of $e^{+}e^{-} \rightarrow \omega \pi^0 $. 
   Data    are from CMD-2\cite{CMD-2:rho}, SND\cite{SND:rho}, and 
              DM2\cite{DM2:rho}. 
Solid line is  fitted curve.  
Dotted lines  represent the contribution of  each amplitude.
}
  \label{fig:res-rho2}
\end{center}
\end{figure}

\begin{figure}
\begin{center}
  \includegraphics[width=7.5cm,angle=0]{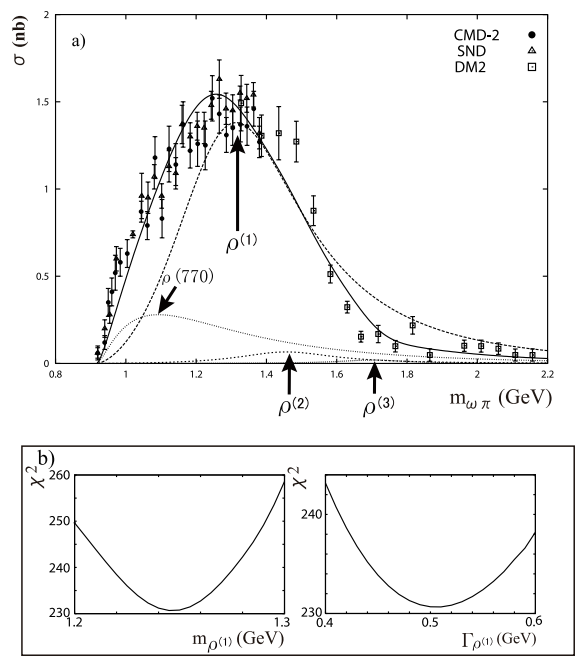}
  \caption{
a) Results of the analysis 
    of 
the 
third case (fit with $ \rho^{(1)} ,\ \rho^{(2)} ,\ \rho^{(3)}$) 
on $\omega \pi^0$ mass spectrum of $e^{+}e^{-} \rightarrow \omega \pi^0 $.
b) Mass and width scan on $\rho^{(1)}$ for 
the 
third case.
   Data are from CMD-2\cite{CMD-2:rho}, SND\cite{SND:rho}, and 
              DM2\cite{DM2:rho}. 
Solid line is  fitted curve.  
Dotted lines  represent the contribution of  each amplitude.
}
  \label{fig:res-rho3}
\end{center}
\end{figure}

 All the results of three cases  are 
shown in Fig. \ref{fig:res-rho1}, Fig. \ref{fig:res-rho2}, 
and Fig. \ref{fig:res-rho3}a).
%-----------------------------------
 In addition mass   and width scan on $\rho^{(1)}$ for 
the 
third case are shown in Fig. \ref{fig:res-rho3}b).
%-----------------------------------
The obtained values of parameters and $\chi^2 / N_{d.o.f.}$ 
 in the three cases  are listed in Table III. 
The result  of fitting of the third case is improved by about $24\sigma$ 
 compared with   the first case, 
indicating the existence of the low mass extra vector meson $\rho'(1.3)$
in addition to two higher state $\rho(1450)$ and $\rho(1700)$. 
%-----------------------------------
%

It may be noted that the CMD-2 data and SND data cover below 1.4 GeV 
of $\omega \pi$ mass spectrum, 
while  
the DM2 data covers above 1.4 GeV. 
The combined data of two regions below and above 1.4 GeV are 
used\footnote{
A bias factor 1.18 is applied on DM2 data by CMD-2.
}
in the present analysis. 
That the DM2 data show their cross section values 
rather  
suppressed would result less contribution of $\rho(1700)$ 
than those of $\rho'(1.3)$ or $\rho(1450)$.

%%%%%%%%%%%%%%%%%%%%%%%%%%%%%%%%%%%%%%%%%
\begin{table}[hbpt]
\begin{center}
\caption{
The obtained values of parameters and $\chi^2 / N_{d.o.f.}$.
}
\begin{tabular}{|c|c|c|c|c|c|c|}
\hline
\multicolumn{7}{|c|}{ Two resonance analysis  } \\   \hline
        &     \multicolumn{3}{|c|}{ ~~~~~~~~~~~~ $\rho^{(1)}$ ~~~~~~~~~~~~ }  &
              \multicolumn{3}{|c|}{ $\rho^{(2)}$ }  \\\hline
Case 1  & \multicolumn{6}{c|}{ 
$\chi^2/N_{d.o.f.} = 801/60 = 13.4 $  
} \\ 
$m(\mbox{\rm MeV})$        &  \multicolumn{3}{|c|}{  1440$^\ddagger$ }  &
                              \multicolumn{3}{|c|}{ 1700 }  \\

$\Gamma (\mbox{\rm MeV})$  &  \multicolumn{3}{|c|}{ $385 \pm  13$ }  
                           &  \multicolumn{3}{|c|}{ 240 }  \\

 A                         &  \multicolumn{3}{|c|}{ $-0.253 \pm 0.009$ }  &
                              \multicolumn{3}{|c|}{ $ 0.008  \pm 0.003$  }  \\ 

\hline % -----------------------------------------------------------------

Case 2  & \multicolumn{6}{c|}{ 
$\chi^2/N_{d.o.f.} = 235/60 = 3.91 $  
} \\ 
$m(\mbox{\rm MeV})$        &  \multicolumn{3}{|c|}{  $ 1251 \pm 11$  }  &
                              \multicolumn{3}{|c|}{ $1700 $  }  \\

$\Gamma (\mbox{\rm MeV})$  &  \multicolumn{3}{|c|}{ $ 563 \pm 19$  }  
                           &  \multicolumn{3}{|c|}{ 240  }  \\

 A                         &  \multicolumn{3}{|c|}{ $-0.545 \pm 0.031 $  }  &
                              \multicolumn{3}{|c|}{ $0.016 \pm 0.003$  }  \\ 

\hline \hline
%%%%%%%%%%%%%%%%%%%%%%%
\multicolumn{7}{|c|}{ Three resonance analysis  } \\   \hline

                           &  \multicolumn{2}{|c|}{ $\rho^{(1)}$  }  &
                              \multicolumn{2}{|c|}{ $\rho^{(2)}$   }  &
                              \multicolumn{2}{|c|}{ $\rho^{(3)}$ }  \\ \hline

Case 3  & \multicolumn{6}{c|}{ 
$\chi^2/N_{d.o.f.} = 231/55 = 4.19$  
} \\ 
$m(\mbox{\rm MeV})$        &  \multicolumn{2}{|c|}{ $1247 \pm 9 $ }  &
                              \multicolumn{2}{|c|}{ 1440$^\ddagger$    }  &
                              \multicolumn{2}{|c|}{ 1700 }  \\ 

$\Gamma (\mbox{\rm MeV})$  &  \multicolumn{2}{|c|}{ $506 \pm 30 $ }  &
                              \multicolumn{2}{|c|}{ 340$^\ddagger$  }  &
                              \multicolumn{2}{|c|}{ 240 }  \\ 

 A                         &  \multicolumn{2}{|c|}{ $-0.435 \pm 0.053 $ }  &
                              \multicolumn{2}{|c|}{  $-0.044 \pm 0.020$  }  &
                              \multicolumn{2}{|c|}{ $0.012 \pm 0.003$  }  \\ \hline
 \multicolumn{7}{l}{ 
$^\dagger$ Bound for upper limit which is set to be consistent 
with PDG tables. } \\
 \multicolumn{7}{l}{ 
$^\ddagger$ Bound for lower limit which is set to be consistent 
with PDG tables. } 
 \end{tabular}
\end{center}
\end{table}
%%%%%%%%%%%%%%%%%%%%%%%%%%%%%%%%%%%%%%%%%%%%%%%%%%%%%%%%%%%%%%%
% ------------------------------------------------------------------------
%-----------------------------------

\newpage
% ---------------------------------------------------------------

%%%%%%%%%%%%%%%%%%%%%%%%%%%%%%%%%%%%%
%%%%%%%%%%%%%%%%%%%%%%

 \ \ \\

\newpage

%%%%%%%%%%%%%%%%%%%%%%%%%%%%%%%%%%%%%
%%%%%%%%%%%%%%%%%%%%%%

%%%%%%%%%%%%%%%%%%%%%%%%%%%%%%%%%%%%%
%%%%%%%%%%%%%%%%%%%%%%

 \ \ \\

\newpage

% --------------------------------------------
\section{Concluding remarks}

Through the present analyses on the mass spectra of $\pi^+\pi^-\pi^0$ 
and of $\omega \pi^0$ 
in $e^+e^-$ annihilation 
we have shown some 
indication of low mass vector mesons  $\omega'(1.3)$ and $\rho'(1.3)$, 
respectively. 
% -------------------------------------
The obtained values of masses and widths are 
\vspace{0.5em}\\
\hspace*{11em} $ m_{\omega'(1.3)} = 
                   1346  \pm  26 \ 
                   (\mbox{\rm MeV}), $ \\
\hspace*{11em} $ ~\Gamma_{\omega'(1.3)} = ~ 
                   639  \pm  42 \ 
                   (\mbox{\rm MeV}),$ \\
\hspace*{11em} $ m_{\rho'(1.3)} = 
                   1247  \pm 9 \ 
                   (\mbox{\rm MeV}), $\\
\hspace*{11em} $ ~\Gamma_{\rho'(1.3)}  = ~ 
                   506  \pm 30 \ 
                   (\mbox{\rm MeV}).$
 \vspace{0.5em} \\
\indent
The results are still preliminary,   
as the used data are combined  one coming from 
different experiments performed in different mass 
regions.
However, we expect that the main 
feature on mass spectra may be 
considered to be 
maintained,  independently   of a bias factor.  
% -------------------------------------

Accordingly, we may conclude that 
the extra-vector mesons $\omega'(1.3)$ and $\rho'(1.3)$ 
are necessary to explain 
the mass spectra.   
% -------------------------------------
In each result of the third case in Table II and Table III, 
the contributions  of 
$\omega(1420)$ and $\rho(1450)$ are very small 
compared with  those  of 
$\omega'(1.3)$ and $\rho'(1.3)$, respectively.
% ---

The present results seem to be consistent with the expectation of  
the $\widetilde U(12)$-sheme. 
In this scheme $\omega'(1.3)$ and $\rho'(1.3)$ are assigned 
as S-wave chiral states\cite{yamada:proc}  while $\omega(1420)$ and $\rho(1450)$ are 
assigned as P-wave states. 
 Accordingly the contributions of  
$\omega(1420)$ and $\rho(1450)$ are expected to be 
very small compared with 
those of  $\omega'(1.3)$ and $\rho'(1.3)$, respectively, 
reflecting the strength  at the origin  of their wave functins, 
$|\psi_{P}(0)|^2 \simeq 0$ and 
$|\psi_{S}(0)|^2 \fallingdotseq 1$. 
% -------------------------------------

Further studies are expected to  confirm 
the existence of 
the low mass extra vector mesons 
$\omega'(1.3)$ and $\rho'(1.3)$: 
%
% -------------------------------------
It 
will be 
 a very important problem for 
hadron spectroscopy.

% -------------------------------------
%%%%%%%%%%%%%%%%%%%%%%%%%%%%%%%%%%%%%%%%%%%%%%%%%%%%%%%%%%%%%%%%%%%%%%
\section*{Acknowledgements}
We would like to express our sincere appreciation to Dr. M. Ishida 
who initiated this works and gave us helpful and crucial suggestions 
during the work.
  We deeply appreciate Prof. K. Yamada, Prof. K. Takamatsu, and Prof. I. Yamauchi
who gave us useful and crucial 
discussions  and information.  
We also thank 
Prof. S. Ishida, Prof. T. Tsuru, Dr. T. Maeda, and Prof. M. Oda 
for useful discussions.  
This work is supported by Nihon University Individual Research Grant for (2005).

% -------------------------------------

\end{document}

%% file: fig-diagram_omega.tex
%WinTpicVersion3.08
\unitlength 0.1in
\begin{picture}( 33.5000, 13.5000)(  5.8000,-16.1000)
% LINE 2 0 3 0
% 2 800 800 1400 1200
% 
\special{pn 8}%
\special{pa 800 800}%
\special{pa 1400 1200}%
\special{fp}%
% LINE 2 0 3 0
% 2 1400 1210 780 1600
% 
\special{pn 8}%
\special{pa 1400 1210}%
\special{pa 780 1600}%
\special{fp}%
% LINE 2 2 3 0
% 2 1410 1210 2020 1210
% 
\special{pn 8}%
\special{pa 1410 1210}%
\special{pa 2020 1210}%
\special{dt 0.045}%
% LINE 2 0 3 0
% 2 2010 1180 2600 1180
% 
\special{pn 8}%
\special{pa 2010 1180}%
\special{pa 2600 1180}%
\special{fp}%
% LINE 2 0 3 0
% 2 2010 1240 2590 1240
% 
\special{pn 8}%
\special{pa 2010 1240}%
\special{pa 2590 1240}%
\special{fp}%
% LINE 2 0 3 0
% 2 2600 1180 3190 810
% 
\special{pn 8}%
\special{pa 2600 1180}%
\special{pa 3190 810}%
\special{fp}%
% LINE 2 0 3 0
% 2 2590 1230 3200 1610
% 
\special{pn 8}%
\special{pa 2590 1230}%
\special{pa 3200 1610}%
\special{fp}%
% STR 2 0 3 0
% 3 580 730 580 830 2 0
% $e^+$
\put(5.8000,-8.3000){\makebox(0,0)[lb]{$e^+$}}%
% STR 2 0 3 0
% 3 590 1570 590 1670 2 0
% $e^-$
\put(5.9000,-16.7000){\makebox(0,0)[lb]{$e^-$}}%
% STR 2 0 3 0
% 3 1710 1380 1710 1480 2 0
% $\gamma$
\put(17.1000,-14.8000){\makebox(0,0)[lb]{$\gamma$}}%
% STR 2 0 3 0
% 3 2220 1390 2220 1490 2 0
% $\omega$
%\put(22.2000,-14.9000){\makebox(0,0)[lb]{$\omega$ , $\phi$}}%
\put(22.2000,-14.9000){\makebox(0,0)[lb]{``$V$"}}%
% STR 2 0 3 0
% 3 3200 790 3200 890 2 0
% $\rho$
\put(32.0000,-8.9000){\makebox(0,0)[lb]{$\rho$}}%
% STR 2 0 3 0
% 3 3240 1600 3240 1700 2 0
% $\pi$
\put(32.4000,-17.0000){\makebox(0,0)[lb]{$\pi$}}%
% LINE 2 0 3 0
% 2 800 800 1400 1200
% 
\special{pn 8}%
\special{pa 800 800}%
\special{pa 1400 1200}%
\special{fp}%
% LINE 2 0 3 0
% 2 1400 1210 780 1600
% 
\special{pn 8}%
\special{pa 1400 1210}%
\special{pa 780 1600}%
\special{fp}%
% LINE 2 2 3 0
% 2 1410 1210 2020 1210
% 
\special{pn 8}%
\special{pa 1410 1210}%
\special{pa 2020 1210}%
\special{dt 0.045}%
% LINE 2 0 3 0
% 2 2010 1180 2600 1180
% 
\special{pn 8}%
\special{pa 2010 1180}%
\special{pa 2600 1180}%
\special{fp}%
% LINE 2 0 3 0
% 2 2010 1240 2590 1240
% 
\special{pn 8}%
\special{pa 2010 1240}%
\special{pa 2590 1240}%
\special{fp}%
% LINE 2 0 3 0
% 2 2600 1180 3190 810
% 
\special{pn 8}%
\special{pa 2600 1180}%
\special{pa 3190 810}%
\special{fp}%
% LINE 2 0 3 0
% 2 2590 1230 3200 1610
% 
\special{pn 8}%
\special{pa 2590 1230}%
\special{pa 3200 1610}%
\special{fp}%
% STR 2 0 3 0
% 3 580 730 580 830 2 0
% $e^+$
\put(5.8000,-8.3000){\makebox(0,0)[lb]{$e^+$}}%
% STR 2 0 3 0
% 3 590 1570 590 1670 2 0
% $e^-$
\put(5.9000,-16.7000){\makebox(0,0)[lb]{$e^-$}}%
% LINE 2 0 3 0
% 2 3320 800 3890 410
% 
\special{pn 8}%
\special{pa 3320 800}%
\special{pa 3890 410}%
\special{fp}%
% LINE 2 0 3 0
% 2 3330 790 3920 1180
% 
\special{pn 8}%
\special{pa 3330 790}%
\special{pa 3920 1180}%
\special{fp}%
% STR 2 0 3 0
% 3 3930 330 3930 430 2 0
% $\pi$
\put(39.3000,-4.3000){\makebox(0,0)[lb]{$\pi$}}%
% STR 2 0 3 0
% 3 3930 1150 3930 1250 2 0
% $\pi$
\put(39.3000,-12.5000){\makebox(0,0)[lb]{$\pi$}}%
\end{picture}%